# Neutron Energy Spectrum Measurements with a Compact Liquid Scintillation Detector on EAST

**Xi Yuan,** [a] **Xing Zhang,** [a] **Xufei Xie,** [a] **G. Gorini,** [b] **Zhongjing Chen,** [a] **Xingyu Peng,** [a] **Jinxiang Chen,** [a] **Guohui Zhang,** [a] **Tieshuan Fan,** [a,*] **Guoqiang Zhong,** [c] **Liqun Hu,** [c] **Baonian Wan,** [c]

[a] *School of Physics and State Key Laboratory of Nuclear Physics and Technology, Peking University, Chengfu Road 201,100871 Beijing, China*

[b] *CNISM and Physics Department, Milano-Bicocca University, 20126 Milano, Italy*

[c] *Institute of Plasma Physics, Chinese Academy of Sciences, PO Box 1126, 230031 Hefei, Anhui, China*

*E-mail:* tsfan@pku.edu.cn

ABSTRACT: A neutron detector based on EJ301 liquid scintillator has been employed at EAST to measure the neutron energy spectrum for D-D fusion plasma. The detector was carefully characterized in different quasi-monoenergetic neutron fields generated by a 4.5 MV Van de Graaff accelerator. In recent experimental campaigns, due to the low neutron yield at EAST, a new shielding device was designed and located as close as possible to the tokamak to enhance the count rate of the spectrometer. The fluence of neutrons and gamma-rays was measured with the liquid neutron spectrometer and was consistent with $^{3}$He proportional counter and NaI (Tl) gamma-ray spectrometer measurements. Plasma ion temperature values were deduced from the neutron spectrum in discharges with lower hybrid wave injection and ion cyclotron resonance heating. Scattered neutron spectra were simulated by the Monte Carlo transport Code, and they were well verified by the pulse height measurements at low energies.

## Contents

# 1 Introduction

Neutron spectrometry is a useful fusion plasma diagnostics. From the analysis of the measured neutron spectrum, various plasma parameters, such as the fuel ion composition, ion velocity distributions and ion temperature, can in principle be determined [1,2]. In fusion experiments, neutrons are mainly produced by fusion reactions between fuel ions, i.e., deuterons (d) or tritons (t). For the case of thermonuclear plasmas, the fuel ions have a Maxwellian velocity distribution and the ion temperature can be directly deduced from the Doppler broadening of the neutron energy spectrum. With auxiliary heating, such as neutral beam injection (NBI) or ion cyclotron resonance heating (ICRH) in the 2$^{nd}$ or 3$^{rd}$ harmonic frequency region of fuel ions, supral-thermal ions are generated and dominate the neutron emission. It is in principle possible to assess the fast ion velocity distribution and evaluate the performance of auxiliary heating from neutron energy spectrum measurements [3,4] and γ–ray spectrum measurements [5,6,7]. The neutron emission spectroscopy (NES) can also be used to study the fast fuel ions destabilizing or stabilizing MHD modes [8].

With this aim, neutron spectrometers have been developed and constructed around tokamak devices. TOFOR [9,10] and MPRu [11,12,13] at JET, which base on the time-of-flight method and magnetic proton recoil method respectively, are the most advanced neutron spectrometers with high energy resolution. Liquid scintillator neutron spectrometers, also referred to as compact neutron spectrometers (CNS) rely on the elastic scattering of fast neutrons in the scintillator material. They have been installed on various tokamak devices, such as JET [14,15], AUG [16], FTU [17], MAST [18], etc. For the neutron energy spectrum measurement at EAST, where the total neutron yield is about $10^9$-$10^{11}$ n/s now, a CNS detector was selected due to its high detection efficiency combined with good $n/\gamma$ discrimination ability and compact size.

In this work, the results of NES measurements with a CNS detector on EAST are described. The CNS detector was calibrated in quasi-monoenergetic neutron fields with neutron energy from 1.1 MeV to 17.5 MeV. After careful characterization, the response matrix of the CNS detector was obtained and used for analysis of measured pulse height spectra of a 2.5 MeV neutron source from d-D reaction. The CNS was set in the EAST experimental hall, with specific shielding for radiation and magnetic field. Time traces of neutrons and γ-rays from the CNS were compared with the results by a $^3$He proportional counter and a NaI (Tl) γ-ray spectrometer. Under different heating scenarios, such as lower hybrid wave injection (LHW) and ICRF, the plasma ion temperature was inferred by searching the best fit between the experimental pulse height spectrum and the theoretical one, which was calculated by folding the theoretical neutron spectrum with the detection response. To assess the ratio of scattered neutrons, a 3-D Monte Carlo model of the EAST device was used to calculate the scattered neutron spectrum. The result was combined with the direct neutron energy spectrum to get the total neutron energy spectrum at the CNS position.

## 2 Calibration of the CNS detector

Calibration of the CNS detector was performed in the neutron fields generated by 4.5 MV Van de Graaff accelerator at Peking University. The EJ301 cylindrical liquid scintillator of the CNS, 5 cm in length and 5 cm in diameter, is optically coupled to a 2 inch Hamamatsu R329-02 photomultiplier tube (PMT) through a cone-shaped light guide. The PMT is coupled to an ORTEC 265A voltage divider and the magnetic field is shielded by an auxiliary 0.8 mm $\mu$-metal.

The liquid scintillation detector was carefully calibrated in order to determine the pulse height function $L(E)$ and the pulse height resolution function $dL/L$, which are different for each individual liquid scintillation detector [19]. As a first step, energy calibration with γ-ray sources was used to obtain the relationship between pulse height and equivalent (Compton) electron energy. For electron energies below 1.6 MeV, this relationship is known to be linear and is determined to be [20]

$$L(E) = G(E_C - 5\text{ keV}) \tag{1}$$

where $E_C$ is the Compton electron energy, and the calibration factor $G$ is expressed in terms of channels per electron energy. Three γ-ray sources $^{137}$Cs, $^{54}$Mn and $^{22}$Na with γ-ray energies of 511keV, 662keV, 835keV, 1275 keV were employed for the calibration. The experimental pulse height spectra were fitted with the theoretical ones calculated by the Monte Carlo code GRESP [21], which was developed at Physikalisch-Technische Bundesanstalt (PTB). After iterative fitting process, the factor $G$ was then determined. The comparison between the pulse height spectrum measured for $^{22}$Na and the theoretical one is shown in Fig.1 (a).

The light output function for protons is nonlinear [19], and it is different for each individual detector. Therefore, it must be carefully determined through the experimental investigation combined with model calculations. The schematic diagram of the experimental setup for the quasi-monoenergetic neutron source calibration is shown in Fig. 2. Pulsed neutron beams with repetition frequencies of 2.5 MHz and 5 MHz were produced by the 4.5 MV Van de Graaff accelerator at Peking University. Two separate methods were employed to discriminate the specific quasi-monoenergetic neutrons from other neutrons and γ-rays, i.e., the time-of-flight measurement and the $n/\gamma$ discrimination technique. For the time-of-flight measurement, the start time and stop time are determined from the anode signal of the CNS detector and the pickup signal of the accelerator, respectively. The $n/\gamma$ discrimination technique is based on the delay-line-shaping method [22]. These two techniques have been effective in rejecting a large fraction of events due to scattered neutrons and γ-rays.

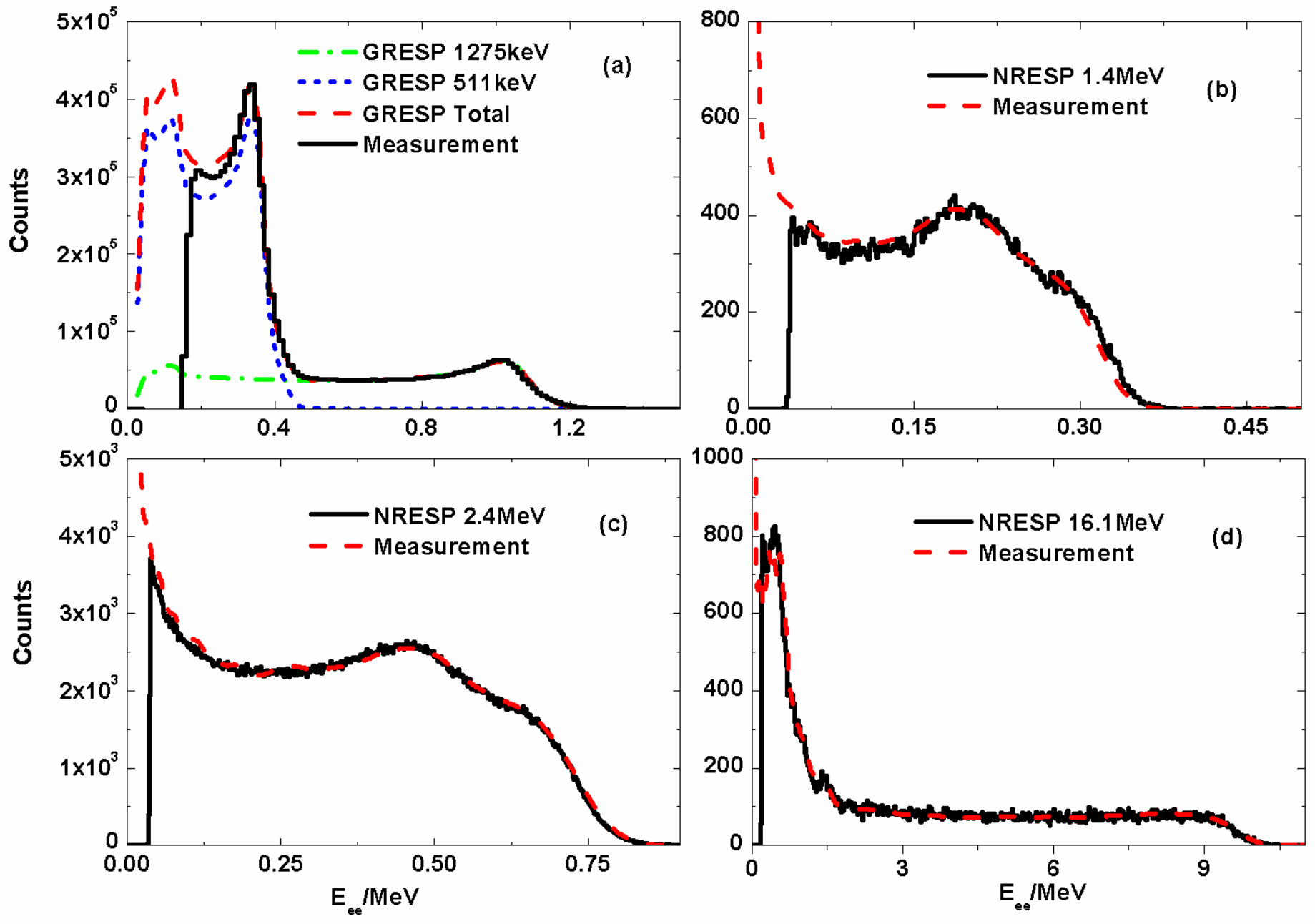

**Figure 1.** Comparison of experimental (black histogram) and theoretical (color dash line) pulse height spectra for $^{22}$Na radiation source (a) and quasi-monoenergetic neutron sources with the energy of 1.4 MeV (b), 2.4 MeV (c) and 16.1 MeV (d). The theoretical pulse height spectra for gamma-rays and neutrons are calculated by GRESP and NRESP, respectively (see text).

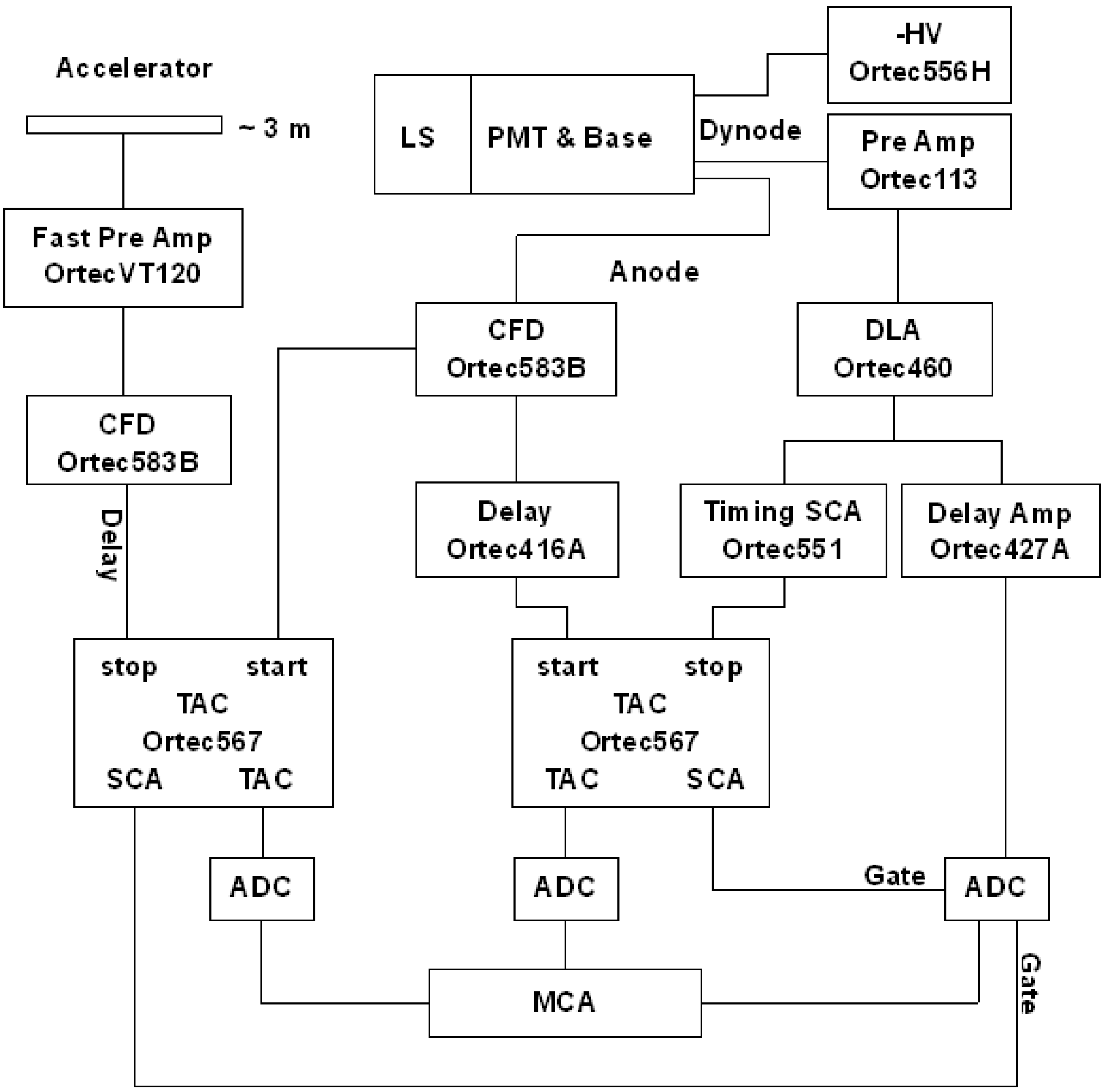

**Figure 2.** Block scheme of the arrangement used to measure the n/γ discrimination by the delay-line-shape method, the time-of-flight spectrum and selected liquid scintillator pulse height spectrum.

The light output of the EJ301 cylindrical liquid scintillator due to monoenergetic protons was obtained by analysis of recoil proton spectra from different monoenergetic (1.1 MeV to 17.7 MeV) neutron injections into the scintillator. The detailed information about the neutron fields is shown in Table 1. Neutrons were generated by the $^{7}$Li(p, n)$^{7}$Be, T(p, n)$^{3}$He, D(d, n)$^{3}$He and T(d, n)$^{4}$He reactions. The energy and broadening of each source were calculated by the TARGET code [23] using the information of ions and targets for the neutron generating reactions. The theoretical spectra were simulated with the Monte Carlo code NRESP [24]. By iterative fitting of calculated spectra to measured spectra, the light output function of protons was finally obtained. Shown in Fig. 1 (b), (c) and (d), are three experimental spectra compared to the calculated spectra. The pulse height resolution function of the CNS detector $dL/L$ was determined using the parameterization [19], which could be expressed as

$$dL/L = \sqrt{\alpha^2 + \beta^2/L + \gamma^2/L^2} \qquad (2)$$

where $L$ is the light output, $\alpha$ is the resolution contribution caused by the position-dependent light transmission from the scintillator to the photomultiplier, $\beta/\sqrt{L}$ is the statistical contribution to resolution due to light production, attenuation, photon to electron conversion and electron amplification in the dynode chain, and $\gamma/L$ expresses all noise contributions to the resolution. For our CNS detector $\alpha = 7.5$, $\beta = 8.08$ and $\gamma = 0.2$.

**Table 1.** Reaction, neutron emission angle, energy and broadening of quasi-monoenergetic neutron fields used for the calibration

| **Num.** | **Reaction** | **Angle (°)** | **$E_n$ (MeV)** | **Broadening (keV)** |
|---|---|---|---|---|
| **1** | $T(d, n)^4He$ | 0 | 17.7 | 767 |
| **2** | $T(d, n)^4He$ | 60 | 16.1 | 426 |
| **3** | $T(d, n)^4He$ | 80 | 15.1 | 260 |
| **4** | $T(d, n)^4He$ | 101 | 14 | 165 |
| **5** | $D(d, n)^3He$ | 0 | 6.1 | 415 |
| **6** | $D(d, n)^3He$ | 0 | 5.2 | 552 |
| **7** | $D(d, n)^3He$ | 66.7 | 4.1 | 155 |
| **8** | $D(d, n)^3He$ | 95 | 2.9 | 73 |
| **9** | $T(p, n)^3He$ | 0 | 2.4 | 200 |
| **10** | $T(p, n)^3He$ | 30 | 2.1 | 180 |
| **11** | $T(p, n)^3He$ | 45 | 1.9 | 160 |
| **12** | $T(p, n)^3He$ | 60 | 1.6 | 140 |
| **13** | $^7Li(p, n)^7Be$ | 0 | 1.4 | 8 |
| **14** | $^7Li(p, n)^7Be$ | 0 | 1.1 | 9 |

As a first assessment of the CNS detector ability to resolve the narrow neutron energy spectra expected for ohmic fusion plasma, the spectrometer was used for neutron energy spectrum measurement in a neutron field generated by deuteron ions impinging on a deuterated target. The CNS detector was located at an angle of 113° relative to the direction of a 3.4 MeV deuteron ion beam. The unfolding code MAXED [25] was applied to unfold the neutron energy spectrum from the experimental pulse height spectrum with the response matrix. The default input energy spectrum in MAXED was set to a flat spectrum, which did not introduce any prior information for the unfolding. The result is shown in Fig. 3. The reconstructed pulse height spectrum by MAXED is generally consistent with the experimental one. Comparison of the neutron energy spectrum unfolded by MAXED and the theoretical one is illustrated in Fig. 3 (b), where the theoretical one is calculated by TARGET. It can be seen that they are generally consistent. Especially, the full width at half maximum (FWHM) of the unfolded neutron spectrum is 50 keV, which is close to the 55 keV calculated by TARGET. This result suggests that the CNS detector can be used for the neutron diagnostic in deuteron plasma discharge experiments. The discrepancies between calculated and measured spectra (e.g. on the low energy side of the peak) are probably due to the limited statistical accuracy of the data and the tendency of MAXED to favor Gaussian spectral shapes.

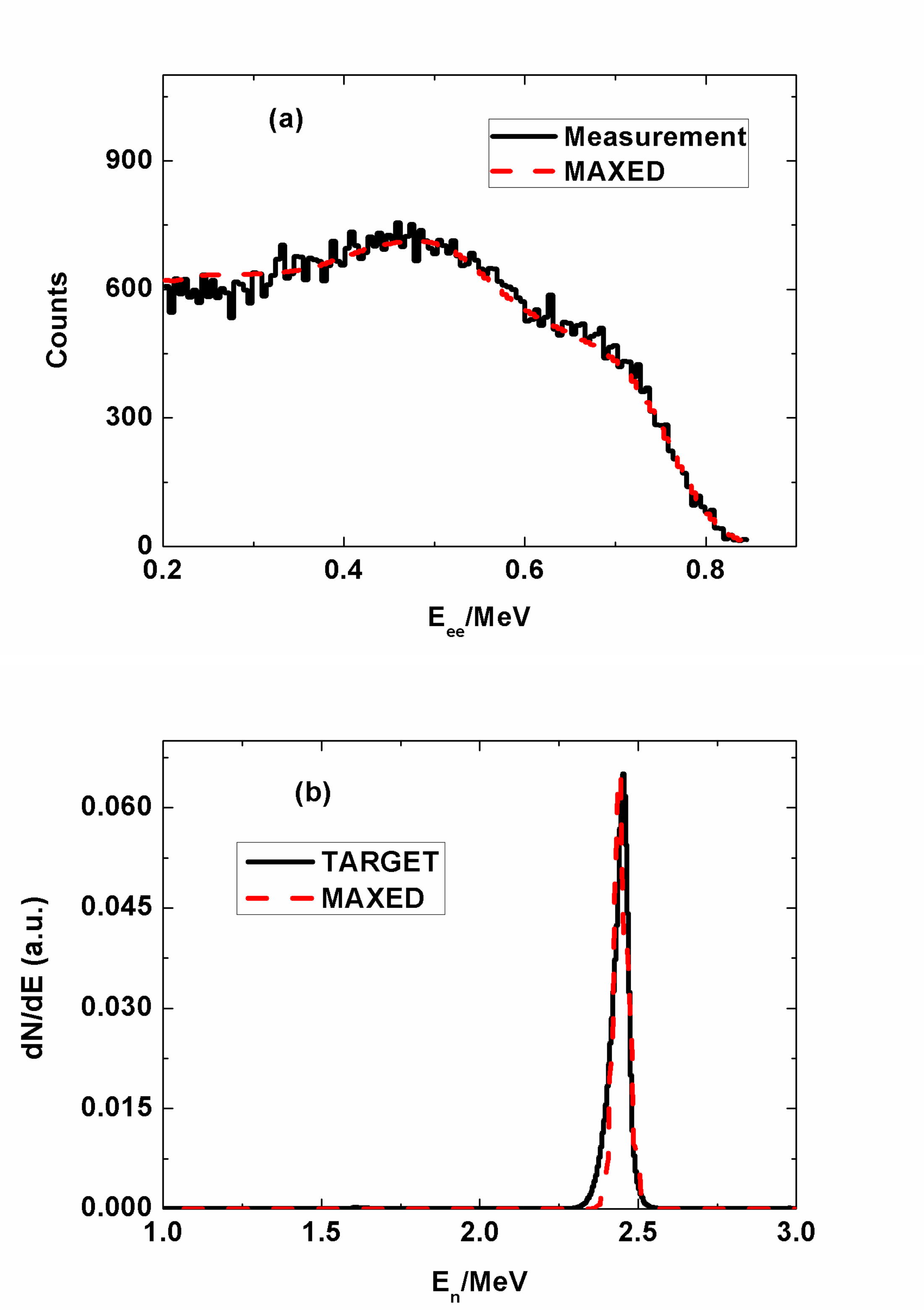

**Figure 3.** (a) Comparison of the reconstructed pulse height spectrum by MAXED and the measured spectrum form the 2.5 MeV neutron source. (b) Comparison of the neutron energy spectrum unfolded with MAXED and the theoretical neutron energy spectrum calculated by TARGET.

## 3 Application of the CNS detector at EAST

EAST is the first non-circular cross-section fully superconducting tokamak in the world. It is constructed for the investigation of some key problems in future fusion devices [26]. Main parameters of EAST are major radius $R$ = 1.85 m, minor radius $a$ = 0.45 m, $I_P$= 1 MA and $B_T$= 3.5 T; the discharge duration could be up to 1000 s. In recent experimental campaigns, the auxiliary heating scenarios include 2 MW lower hybrid wave injection (LHW) and 4.5 MW ion cyclotron resonance frequency heating (ICRH).

The CNS was installed inside the experimental hall（see Fig. 4）, at the mid-plane of the plasma and about 3.5 m away from the plasma core, as close as possible to the plasma in order to enhance the neutron count rate in view of the relatively low neutron yield. A dedicated shielding composed of 25 cm thick polythene and 10 cm lead was constructed to attenuate scattered neutrons and γ-rays. Around the spectrometer a 1.3 cm DT4C soft iron cylinder and 2 mm permalloy was mounted to protect the detector from the magnetic field, which was less than 0.15 T at the detector position.

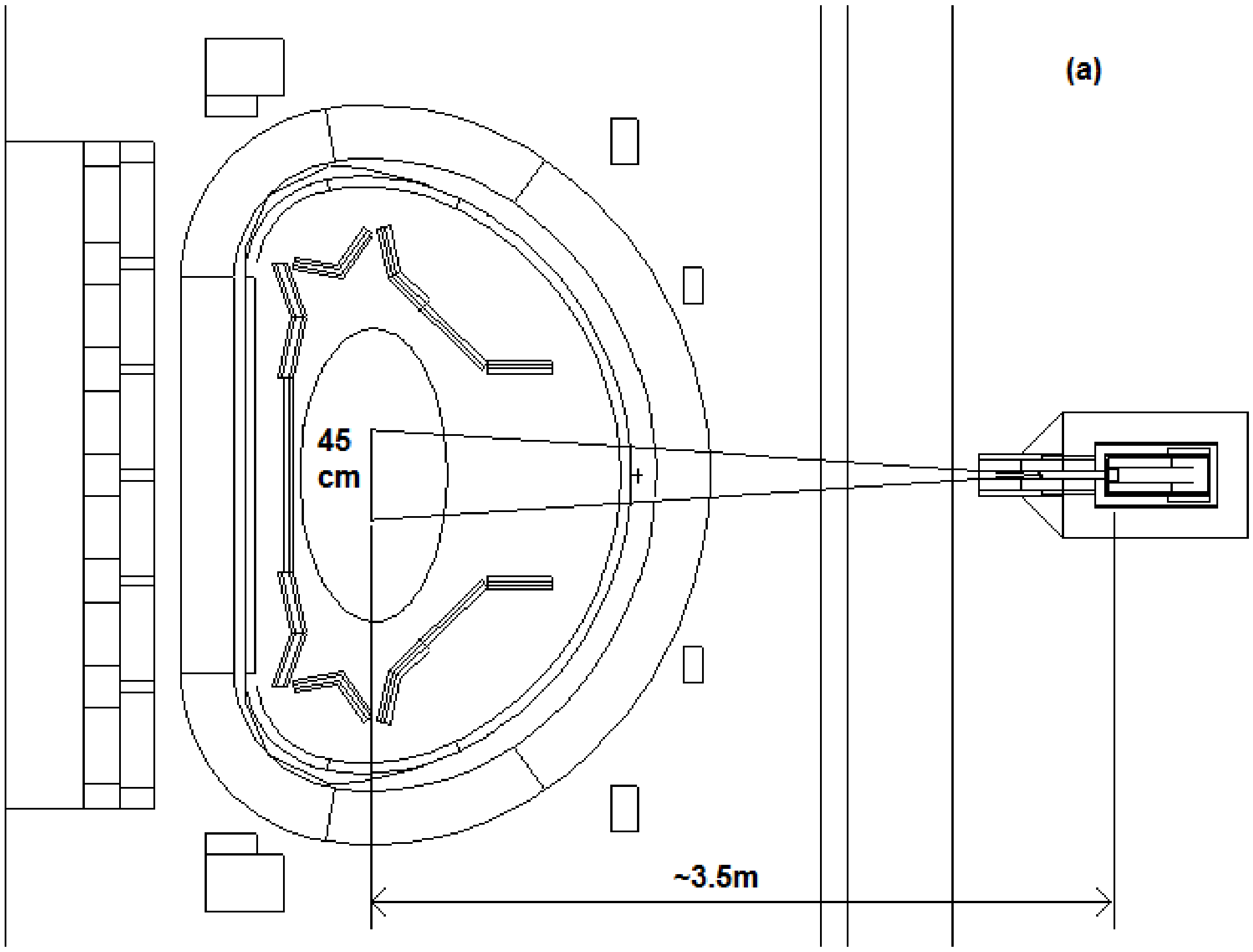

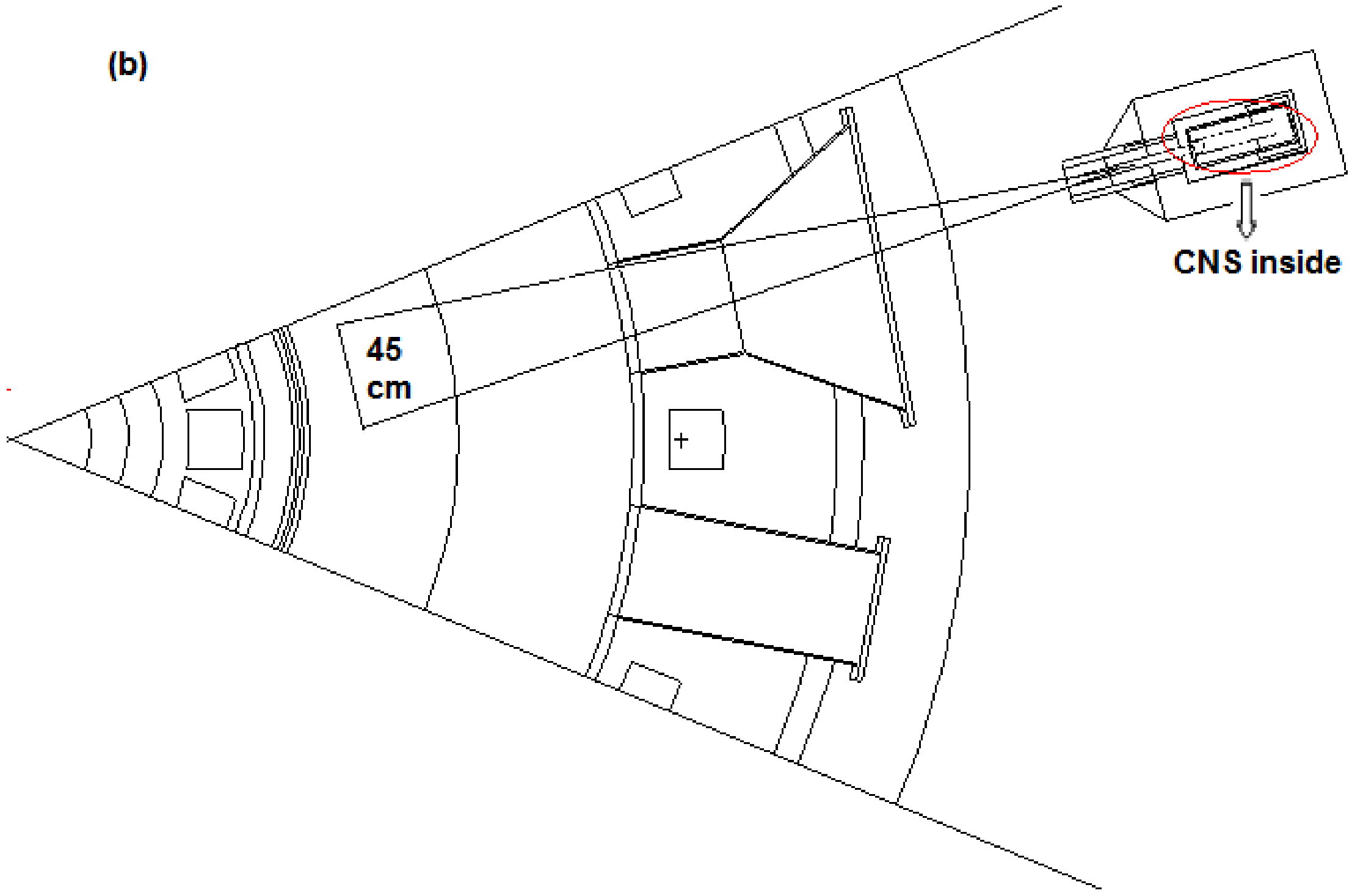

**Figure 4**. Schematic model of the experimental setup of the CNS at EAST with the elevation view (a) and the plane view (b). The spectrometer is placed in a specified shielding indicated in the figure. The model was used for MCNP calculation of the scattered neutron spectrum (see text).

Time traces of neutrons and γ-rays measured by CNS for EAST discharge #40426 are shown in Fig. 5 together with other plasma discharge parameters, namely plasma current $I_p$, electron density $n_e$, LHW power, ICRF power, neutron yield monitored by a $^3$He proportional counter, and γ-ray yield monitored by a NaI(Tl) γ-ray spectrometer. It can be seen that the neutron signal evolution measured by the CNS detector is reasonably consistent with the one measured by the $^3$He proportional counter, and both of them increase with the ICRF heating. Note that γ-ray rates measured by the CNS detector and NaI(Tl) spectrometer are both suppressed by the LHW injection in the plasma.

The performance of the CNS detector was investigated under the different heating scenarios of LHW injection and ICRH available on EAST at the time. In the LHW injection scenario, the wave power was absorbed by plasma electrons. While in the ICRF heating scenario, the heating power was deposited to the minority hydrogen (H) ions due to the low $n_H/n_D$ ratio [27]. The experimental pulse height spectra under different heating scenarios are illustrated in Fig. 6 (a) and (c). In Fig. 6 (a) the measured pulse height spectrum under LHW injection with the power of 1.1 MW is shown. It should be noted that results from thirteen similar discharges (#40978 - #40990) are added up. In Fig. 6 (c) the measured pulse height spectrum from twelve similar discharges (#41110 - #41115 & #41118 - #41122) under ICRF and LHW heating is shown. In both scenarios there are very few events in the high energy region (i.e. $E_{ee}$>0.9 MeV) of the pulse height spectrum, which means that high energy neutrons are rare indicating that very little power is absorbed by fuel deuterium (D) ions directly. Therefore, the velocity distribution function of D ions is Maxwellian, and the neutron spectrum can be represented by a Gaussian with

$$\mathrm{FWHM}(\mathrm{keV}) = 82.5\sqrt{T_i(\mathrm{keV})} \tag{3}$$

In order to estimate the ion temperature ($T_i$), different pulse height spectra were generated by folding the response matrix with Gaussian neutron spectra corresponding to different ion temperatures. The fitting region was limited to $E_{ee} > 0.55MeV$ (which is marked in Fig. 6 by a solid line) in order to limit the effect of scattered neutron on the detector. Then the neutron energy spectra, as shown in Fig. 6 (b) and (d), respectively, were fitted using the $\chi^2$-square method. Care was taken to bin the data on the high energy side until the number of count in each bin was sufficient for applicability of $\chi^2$-square method. The fitted pulse height spectra were generally consistent with the measured ones, as illustrated in Fig. 6 (a) and (c). The ion temperature under LHW injection was about 0.7 keV (with confidence limit from 0.2 to 1.2 keV) and the ion temperature under ICRF heating was about 1.1keV (with confidence limit from 0.3 to 2.0 keV). For comparison the ion temperatures derived from the imaging X-ray crystal spectrometer (XCS) diagnosis were 0.7 keV and 1.0 keV, respectively [28]. The ion temperature from CNS detector and XCS are shown in Fig. 7.

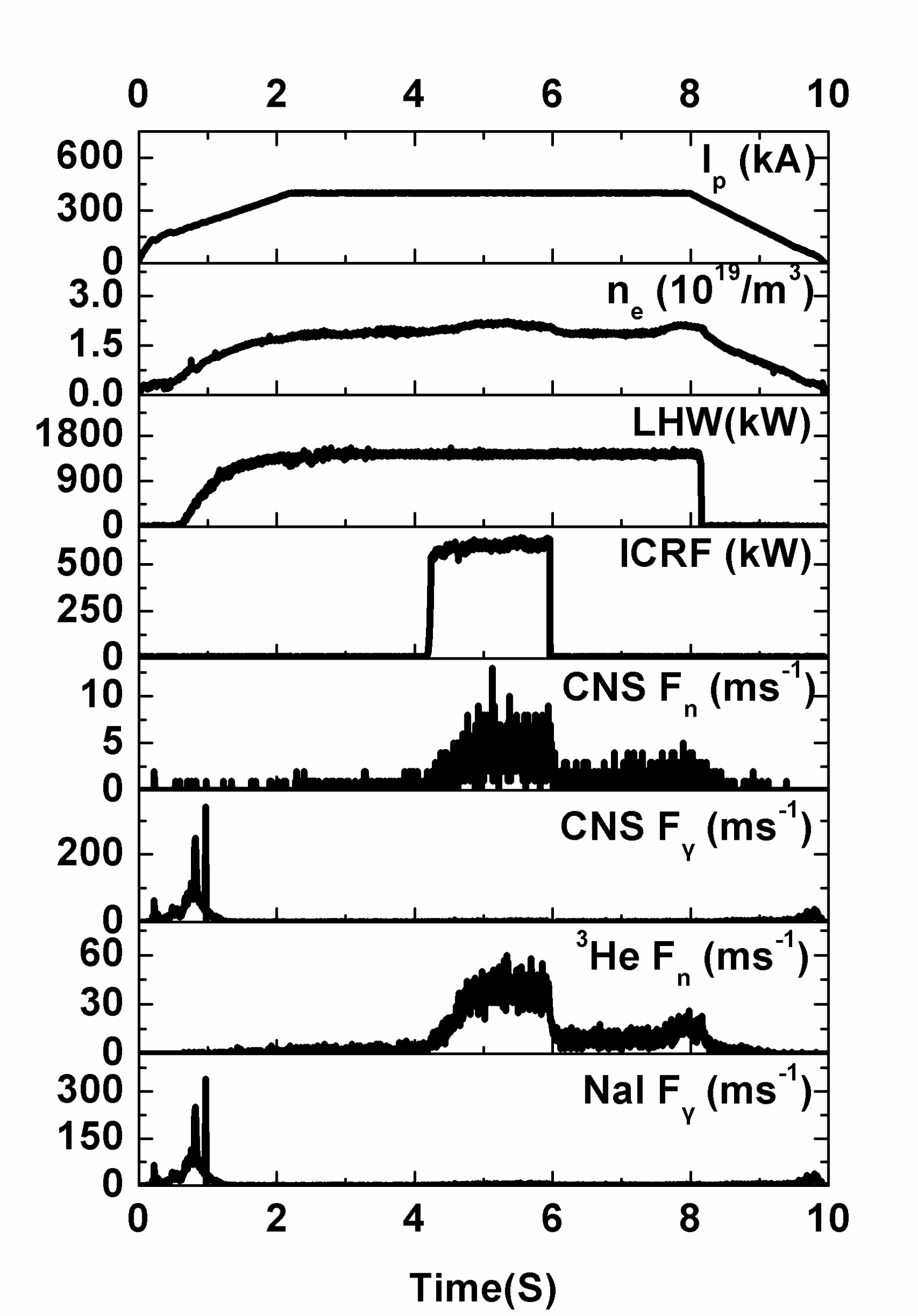

**Figure 5**. Time traces of plasma current, electron density, LHW power, NBI power, neutron and γ-ray count rate from CNS and from other two diagnostics during EAST plasma discharge #40426.

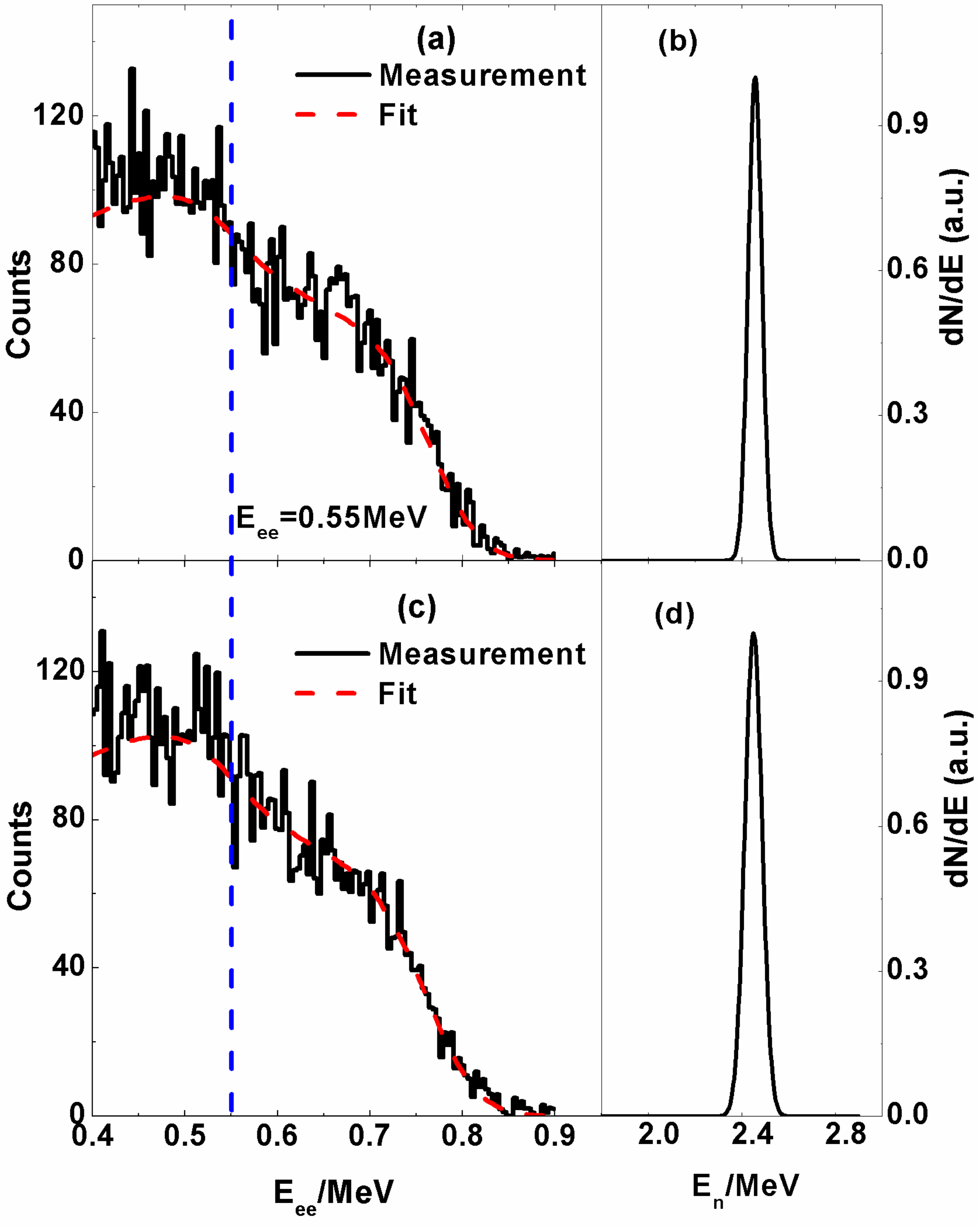

**Figure 6.** Measured pulse height spectra (black histogram) and calculated ones (red dashed line) that result from folding the CNS response to the neutron spectrum on the right side for two scenarios: (a)&(b) for LHW and (c)&(d) for ICRF + LHW.

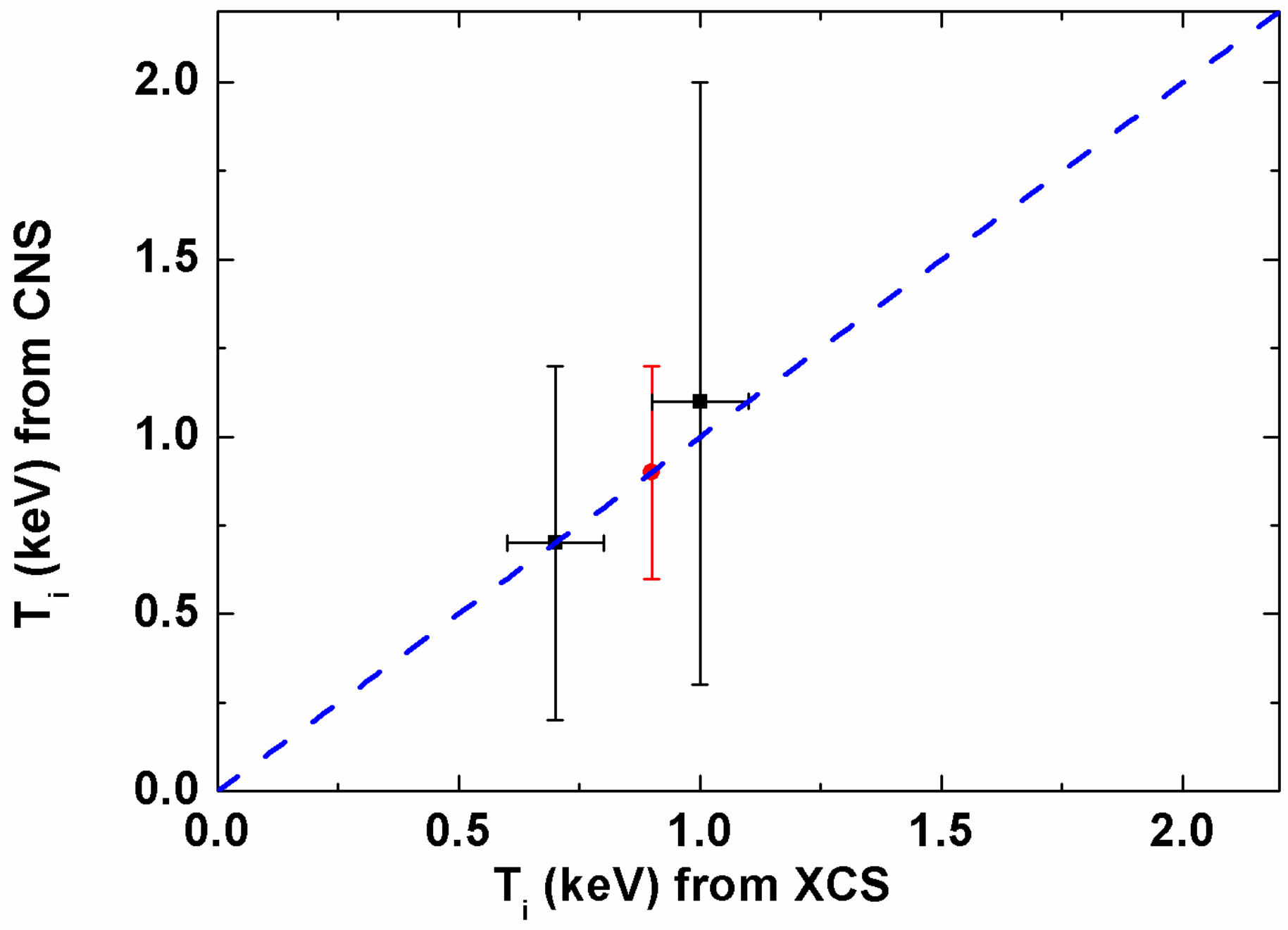

**Figure 7**. Ion temperature from the CNS compared with the ion temperature from the XCS. The value of 0.9 ±0.3 keV (in red) refers to a set of 300 EAST discharges (see text); the corresponding average $T_{\rm i}$ value from XCS was not available and $T_{\rm i}$ = 0.9 keV is used.

The amount of scattered (energy degraded) neutrons reaching the CNS detector was also assessed by the general Monte Carlo N-particle transport Code MCNP [29]. For the analysis of the experimental data in discharge #41110, as shown in Fig.8 (red dashed line), the measured pulse height spectrum at the low energies is well above the one calculated by folding the detector response with a Gaussian neutron spectrum. The discrepancy is due to the scattered neutrons generated in the tokamak device. A detailed MCNP 3-D model was setup with the assistance of the MCAM code [30]. This model, as illustrated in Fig. 2, is 1/8 sector of the tokamak device with reflecting boundary, and it takes into consideration all significant components such as central solenoid, toroidal and poloidal field coils, first wall, thermally conductive layer, diverter, vacuum vessel and so on. The CNS detector and shielding were also modeled. Neutrons were generated in the plasma region and the neutron energy was sampled at DD fusion reactions with the ion temperature of 1.1 keV. The direct neutron energy spectrum and the MCNP simulated energy spectrum are shown in Fig. 9. Note that the peak at $E_n \approx 1.8 MeV$ is due to neutron backscattering off carbon, which is the main component of the first wall. Taking into account the scattered neutrons, the neutron energy spectrum was folded with the response matrix and the resulting pulse height spectrum was compared with the experimental one, as shown in Fig. 8 with a green line. The result shows that the agreement between data and simulation is substantially improved at low pulse heights.

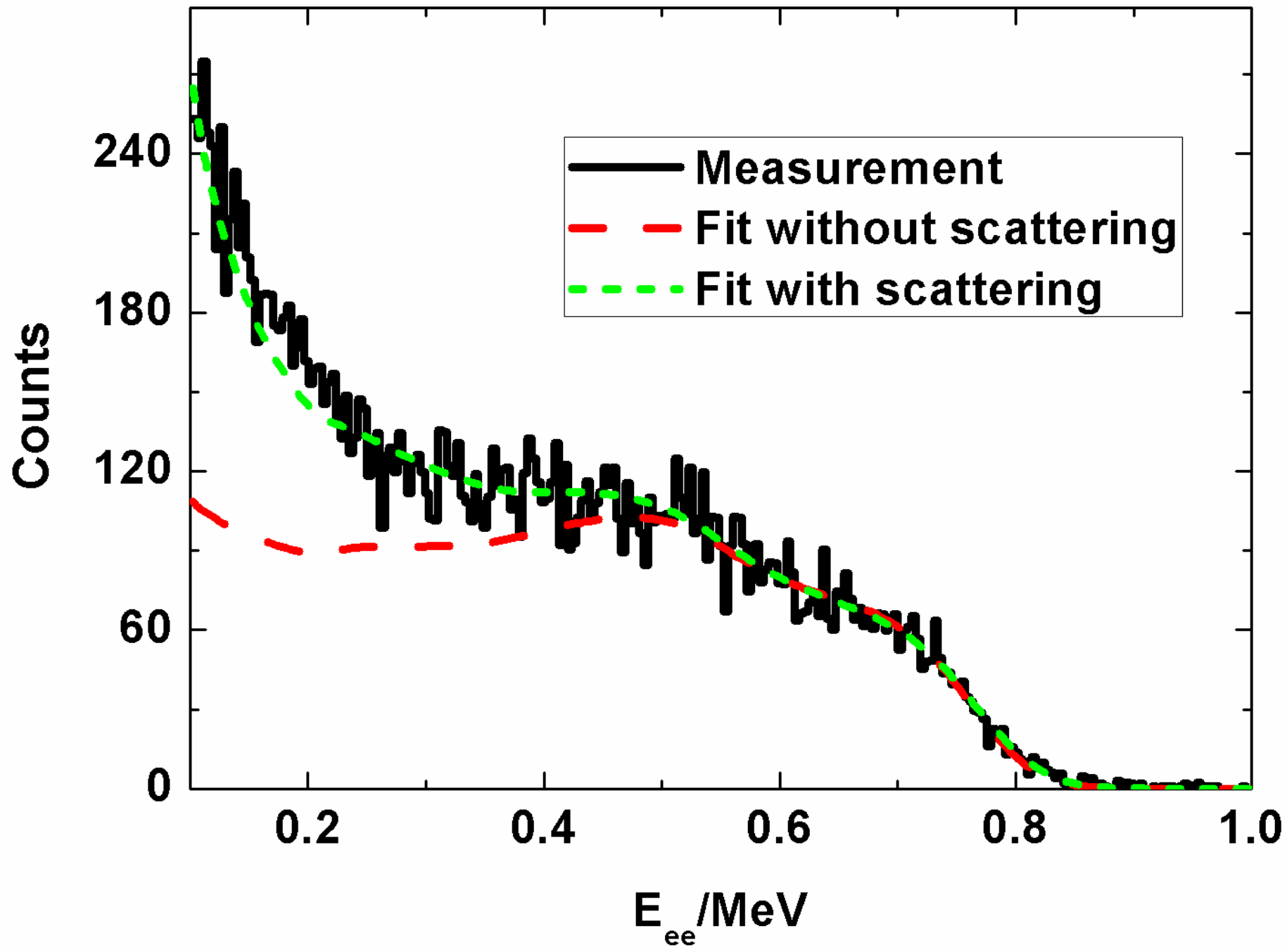

**Figure 8**. The measured pulse height spectrum (black histogram) compared with pulse height spectrum from folding neutron spectrum with (green dash line) and without (read dashed line) scattered neutrons.

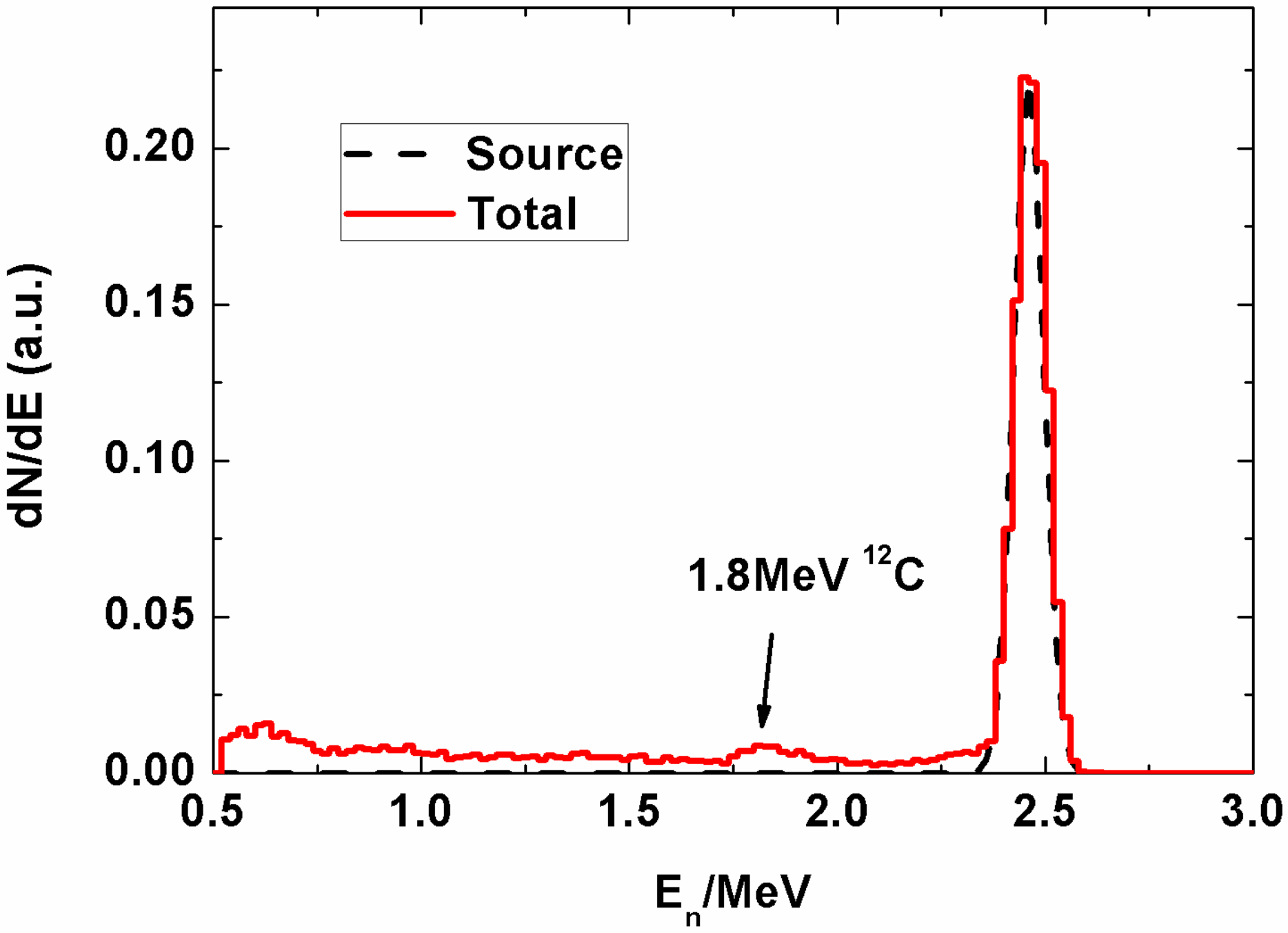

**Figure 9**. The neutron spectrum generated by the DD fusion source (dashed) compared with the neutron spectrum calculated by MCNP (full line).

## 4 Discussion

Generally, in a situation where $T_i$ < 3 keV, NES should not be expected to provide good quality $T_i$ values. A better way to estimate the plasma ion temperature is with the measurement of the total neutron yield, taking advantage of the strong $T_i$ dependence of the neutron emissivity $\eta$ (nearly proportional to $T_i^5$). If one wants to provide useful $T_i$ data from NES, the energy resolution of the spectrometer would have to be very good. However, as the total neutron yield is low at this ion temperature, high resolution neutron energy spectrometers based on time-of-flight method or magnetic proton recoil method would suffer severe problems induced by statistics. Only a spectrometer based on liquid scintillator could provide data with sufficient statistical accuracy. Moreover, it would be expected that the modest resolution of the liquid scintillation detector would be not a fundamental limit to the accuracy of $T_i$ provided all the following requirements are fulfilled (which is the case of our experiment) [31]:

(i) The response matrix of the detector is well known. The detector has been well characterized by quasi-monoenergetic neutron sources and the accuracy of the response matrix has been validated by the measurement of the 2.45 MeV quasi-monoenergetic neutron source.

(ii) The fluence of the scattered neutrons is well understood. With the simulation based on MCNP model, the influence of the scattered neutrons has been well assessed and it is relatively small compared to the direct neutrons.

(iii) The response of the detector is very stable. The detector has been steadily operated over five weeks. The reliability and reproducibility is demonstrated in Fig.10, where the measured data with total 300 discharges has been categorized into three groups according to the experiment date. It could be seen that all pulse height spectra are practically identical within statistics.

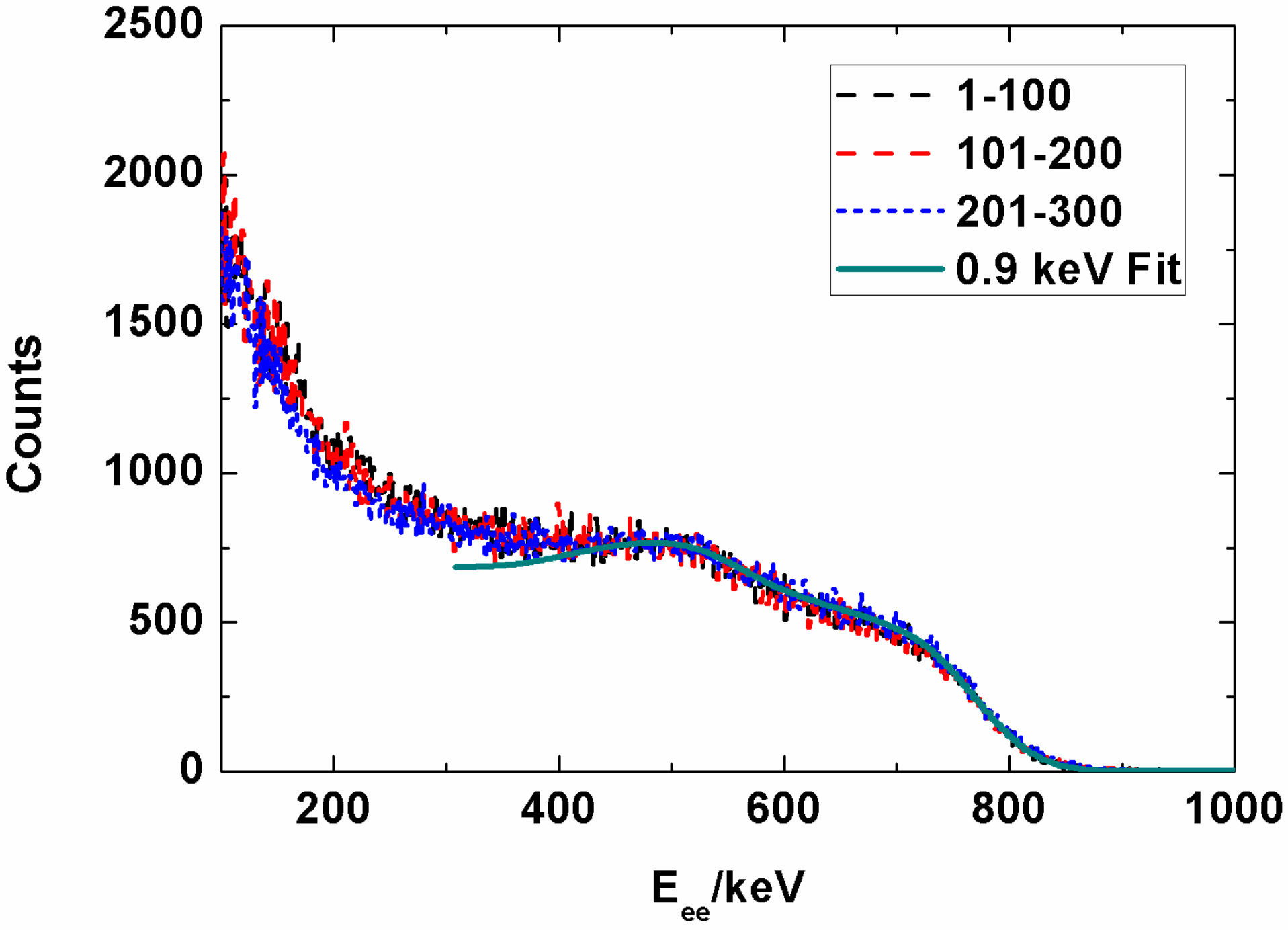

**Figure 10**. Comparison of 3 pulse height spectra groups with 100 discharges (dashed lines) in each one and calculated one (full line) results from folding the CNS response to the Gaussian neutron spectrum with $T_i$ = 0.9keV.

(iv) The influence of the background radiation is low. The detector has been placed in a well designed shielding and the background radiation is effectively suppressed.

If we add together all 300 pulse height spectra shown in Fig.10, we get a fitted $T_\mathrm{i}$ value of about 0.9±0.3 keV. The calculated pulse height spectrum of Gaussian neutron spectrum with the fitted $T_i$ = 0.9keV is also shown in Fig.10.

With detailed analysis of the fluence of scattered neutrons, the CNS detector with C&M system can be used in a neutron camera with added spectrometry function. This could provide, as a minimum, the possibility for good separation of direct and scattered neutrons, especially in edge channels where the scattered neutron flux is > 50%. Nowadays liquid scintillation detectors are in regular use in neutron cameras (JET [32]) and are a prime candidate for application in new devices with high neutron yield (e.g. ITER [33]) as well as low neutron yield (e.g. HL-2A).

In the near future a 2 MW NBI systems will be installed on EAST, and the neutron yield is expected to increase up to $10^{14}$/s. With this high neutron yield, more advanced NES measurements will become possible. A neutron energy spectrometer based on time-of-flight is being designed. Once installed, it will allow for detailed studies of the fast ion behavior in NBI heating scenarios.

## 5 Conclusions

A so-called compact neutron spectrometer based on the EJ301 liquid scintillator has been successfully operated on the EAST tokamak. The CNS has been well characterized at accelerator lab and the detector response was measured. For deuteron discharge experiments at EAST, the neutron spectrum was successfully measured at the low neutron yields and the ion temperature was obtained from the FWHM of the Gaussian neutron peak. A MCNP model was set up to assess the scattered neutrons, which provided a good match to the low energy part of the CNS pulse height spectrum. The experience gained with the CNS on EAST can be used to address the use of the CNS as detector in a camera system with added spectrometry function. Future plans for EAST include the design and installation of a more advanced NES system based on the TOF method.

## Acknowledgments

This work was supported by the state Key Development Program for Basic Research of China (Nos. 2013GB106004, 2012GB101003 and 2008CB717803) and the National Natural Science Foundation of China (No. 91226102). The authors are very grateful to the EAST operation Team for their help during experiment on EAST. The authors are also grateful to Prof. R. Nolte and Dr. M. Reginatto of PTB for the help in using the NRESP，GRESP, MAXED and Target codes. The authors are grateful to Prof. Jianyong Wang of Peking University for the accelerator operation.